# Rate Splitting for MIMO Wireless Networks: A Promising PHY-Layer Strategy for LTE Evolution

**Bruno Clerckx[1,2], Hamdi Joudeh[1], Chenxi Hao[1,3], Mingbo Dai[1] and Borzoo Rassouli[1]**
[1]Imperial College London, [2]Korea University, [3]Beijing Samsung Telecom R&D Center
Email: {b.clerckx,hamdi.joudeh10,chenxi.hao10,m.dai13,b.rassouli12}@imperial.ac.uk

**Abstract**
MIMO processing plays a central part towards the recent increase in spectral and energy efficiencies of wireless networks. MIMO has grown beyond the original point-to-point channel and nowadays refers to a diverse range of centralized and distributed deployments. The fundamental bottleneck towards enormous spectral and energy efficiency benefits in multiuser MIMO networks lies in a huge demand for accurate channel state information at the transmitter (CSIT). This has become increasingly difficult to satisfy due to the increasing number of antennas and access points in next generation wireless networks relying on dense heterogeneous networks and transmitters equipped with a large number of antennas. CSIT inaccuracy results in a multi-user interference problem that is the primary bottleneck of MIMO wireless networks. Looking backward, the problem has been to strive to apply techniques designed for perfect CSIT to scenarios with imperfect CSIT. In this paper, we depart from this conventional approach and introduce the readers to a promising strategy based on rate-splitting. Rate-splitting relies on the transmission of common and private messages and is shown to provide significant benefits in terms of spectral and energy efficiencies, reliability and CSI feedback overhead reduction over conventional strategies used in LTE-A and exclusively relying on private message transmissions. Open problems, impact on standard specifications and operational challenges are also discussed.

## 1. Introduction

Promising approaches for 5G consist in densifying the network by adding more antennas in a distributed or co-localized manner. A distributed deployment leads to dense homogeneous/heterogeneous networks where the widely recognized bottleneck is interference. Interference management relying on multi-point cooperation have drawn a lot of attention in industry (i.e. CoMP in LTE-A [1]) and academia. Co-localized deployment leads to massive MIMO (i.e. FD-MIMO in LTE-A).

Although appealing in their concept, those aforementioned MIMO techniques are hampered by several practical factors. Among these, the acquisition of accurate CSI knowledge at the transmitter (CSIT) is the major challenge. The availability of accurate CSIT is crucial for Downlink (DL) multi-user MIMO wireless networks. The beamforming and interference nulling performance heavily depends on the channel estimation accuracy. Unfortunately, pilot reuse tends to impair channel estimation in TDD and a significant feedback overhead is required to guarantee sufficient feedback accuracy in FDD due to the large number of antennas. Delay and inaccurate calibrations of the RF chains also contribute to making the CSIT inaccurate. CSIT inaccuracy results in a multi-user interference and link adaptation problem that is the primary bottleneck of MIMO wireless networks, as highlighted e.g. in [2] for MU-MIMO and [3] for CoMP.

Looking backward, the problem has been to strive to apply techniques designed for perfect CSIT to scenarios with imperfect CSIT. Following the same path will only increase the gap between theory and practice as the density of antennas increases. The motivation behind this paper is the following: would it not be wiser to design wireless networks from scratch accounting for imperfect CSIT and its resulting multi-user interference?

Interestingly, there has been some recent communication and information theoretic progress in understanding the fundamental impact of imperfect CSIT and resulting multi-user interference on the performance (measured in terms of degree of freedom) of wireless networks. Results highlight that to benefit from imperfect CSIT and tackle the multi-user interference, the transmitter should take a rate-splitting (RS) approach that splits each message into a common and a private message, and superpose a common message on top of all users' private messages. The common message is encoded using a codebook shared by all receivers and is intended to a subset of the users but is decodable by all users, while the private part is to be decoded by the corresponding receiver only. This contrasts with LTE-A MU-MIMO/CoMP/HetNet that are entirely designed based on private messages transmission!

The paper provides a survey on recent advances in RS for MIMO wireless networks in various scenarios such as MU-MIMO, Massive MIMO, Multi-Cell Coordination and highlights its potential and benefits over traditional approaches used in LTE-A. It also identifies the challenges and the necessary standardization efforts to make RS a reality in LTE Evolution.

## 2. Fundamental of Rate Splitting

The concept of RS is not particularly new. Its roots date back to the early works on the two-user Interference channel (IC) by Carleial and Han and Kobayashi [4]. Those authors developed transmission strategies based on RS to achieve new rate regions. In the Han-Kobayashi scheme, which achieves the best known inner bound to date, each source divides its message into a "private" part and a "common" part (sometimes referred to as a "public" part). The two parts are encoded using superposition coding and simultaneously transmitted. In addition to decoding its own message consisting of two parts, each receiver also decodes part of the interference, specifically the other receiver's common part. The beauty of this scheme lies in the fact that it generalizes two extreme strategies: treating interference as noise, and interference decoding. The Han-Kobayashi scheme reduces to one of the aforementioned strategies under extreme conditions, and provides a tradeoff for intermediate regimes.

For the MIMO Broadcast Channel (BC), i.e. the information theoretic counterpart of a single-cell MU-MIMO system, it is well established that the capacity region is achieved using Dirty Paper Coding (DPC) under perfect CSIT. However, DPC is merely a theoretical concept, and its practical implementation is deemed highly complex. Linear precoding strategies have emerged as the most attractive alternative, due to their considerably simpler implementation, and their optimality from a Degrees of Freedom (DoF) point of view. The DoF can be interpreted as the number of interference-free data streams that can be simultaneously communicated per one channel use. This is quantified at very high Signal to Noise Ratios (SNRs), where the effect of additive noise can be neglected and the limiting factor becomes the inter-user interference. The optimality of linear precoding in this sense stems from the fact that it can be utilized to place each user's signal in the null space of all other users e.g. by employing Zero-Forcing beamforming (ZFBF). However, imperfect CSIT knowledge results in distorted interference-nulling and yielding residual interference at the receivers, which in turn may jeopardize the achievable DoF. This draws strong resemblance to the IC, one that has been generally overlooked. Conventionally, transmission strategies have been developed for MU-MIMO systems with imperfect CSIT by treating the residual interference as noise. However, the lessons learnt from the IC and the Han-Kobayashi scheme suggest that it is advisable to decode part of the interference (or the whole of it) under certain circumstances. This motivates the employment of the RS transmission strategy for MU-MIMO systems with imperfect CSIT [5].

The RS strategy for MU-MIMO is formally described as follows. Let $W_1, W_2, \cdots, W_K$ be the uncoded messages intended to users $1,2,\cdots,K$ respectively, simultaneously served by the BS in the same time-frequency resource block. Generally speaking, each message is split into two parts, e.g. $W_{k0}$ and $W_{k1}$ which correspond to the common part and private part of $W_k$ respectively. The ratios in which messages are divided are design parameters that vary depending on the setup. All common parts are packed into one *super* common message, i.e. $W_0 = (W_{10}, \cdots, W_{K0})$. In a linearly precoded system, the resulting *K+1* messages are first encoded into symbol streams, the *K* private streams are then mapped to the transmit antenna array through legacy MU-precoders (e.g. ZFBF), while the common stream is precoded in a multicast fashion such that it is delivered to all users. Each UE performs joint decoding of the common stream plus its private stream. This can be implemented through decoding the common stream first by treating all private streams as noise, followed by decoding the private stream after removing the common stream via Successive Interference Cancellation (SIC). The overall architecture is illustrated in Figure 1.

It should be noted that while the RS transmit signal model resembles a broadcasting system with unicast (private) streams and a multicast stream, the role of the common message is fundamentally different. The common message in a unicast-multicast system carries public information intended as a whole to all users in the system, while the *super* common message in RS encapsulates parts of private messages, and is not entirely required as by all users, although decoded by them all for interference mitigation purposes.

Employing the combination of superposition coding and SIC draws the comparison with Non-Orthogonal Multiple Access (NOMA), also called Multi-User Superposition Transmission (MUST) in LTE Rel-13, recently investigated as a potential strategy for 5G. While the two methods are differently motivated, links between them can be established. For instance, the common message in the RS scheme can be seen as a non-orthogonal layer added onto the conventional orthogonal ZFBF layers. However, generally speaking, the two strategies cannot be treated as extensions or subsets of each other, at least in their currently proposed forms.

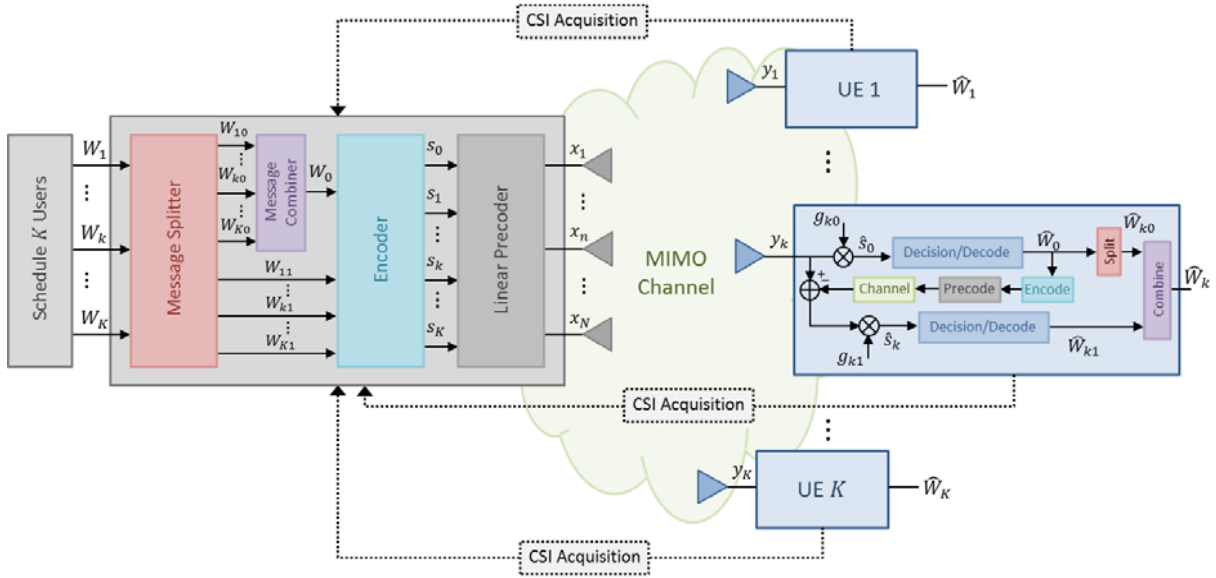

Figure 1. Overall Architecture of MU-MIMO with Rate-Splitting

### 3. Performance Limits and Degrees of Freedom

In DoF analysis of MIMO systems, the CSIT quality is commonly quantified in terms of a non-negative constant exponent $\alpha$ such that errors decay with increased SNR at a rate of $O(\mathrm{SNR}^{-\alpha})$. In limited feedback systems where UEs send quantized versions of their channels back to the BS, $\alpha$ is interpreted in terms of the number of feedback bit. For example, $\alpha = 0$ corresponds to non-scaling scenarios where the number of feedback bits is fixed with SNR, and $\alpha > 0$ corresponds to scenarios where the number of feedback bits scales with SNR. In LTE-A, the number of feedback bits does not scale with the SNR and $\alpha = 0$ is applicable. It is worth highlighting that $\alpha$ can also assume a rather different interpretation. In particular, $\alpha$ can be written in terms of the normalized Doppler frequency in systems where CSIT is somehow outdated, where smaller $\alpha$ represent higher Doppler frequencies.

It is well established that under imperfect CSIT, the maximum DoF of the MIMO-BC can be maintained as long as $\alpha \geq 1$, i.e. CSIT errors decay with SNR at a rate not slower than $O(\mathrm{SNR}^{-1})$. Using ZFBF over the imperfect channel estimate at the BS yields non-dominant residual interference, sufficiently treated as noise from a DoF perspective. However, maintaining such *high* CSIT qualities may be exhausting in terms of resources, and it is not uncommon in practical systems to have $\alpha < 1$ (e.g. with quantized CSI in LTE-A). In such situations, treating the residual multi-user interference as noise is known to deteriorate the DoF performance. For example, in a system where each user is equipped with a single antenna, transmitting data streams along ZFBF vectors achieves a fraction $\alpha$ of the maximum DoF obtained under perfect CSIT, i.e. $K\alpha$. On the other hand, superior DoF performances are achieved when the RS strategy is leveraged. Specifically, decoding part of the interference presented in the form of a common message achieves an extra DoF of $1-\alpha$. However, realizing such gains requires a careful power allocation among the private and common streams; one that guarantees the common stream's DoF gain while not compromising the private streams' achievable DoF. The DoF gains are illustrated in Figure 2.

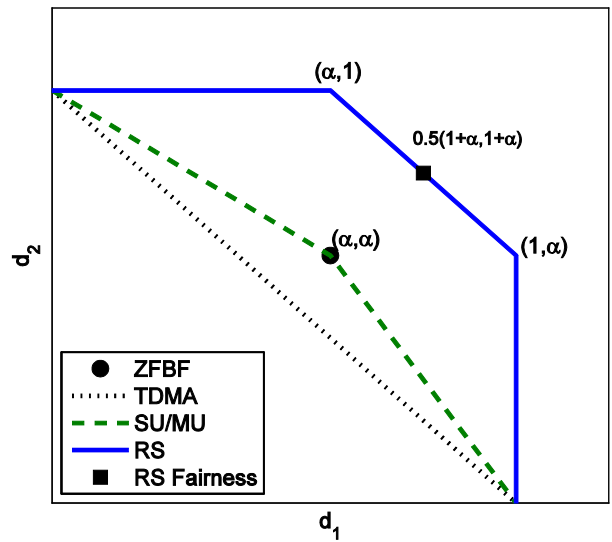

Figure 2. DoF region achievable with rate-splitting and conventional strategies (ZFBF, SU/MU, TDMA) for $\alpha = 0.6$

## 4. Sum-Rate Enhancement and CSI Feedback Reduction

So far the focus has mainly been on the DoF analysis, leaving aside the question of how the Rate-Splitting approach can benefit the sum rate performance at finite SNR. Tackling such a question is essential as it sheds light on the usefulness of the information-theoretic works in practical multiuser MISO systems.

Considering that there are two co-scheduled users and the quantized CSIT is obtained via Random Vector Quantization (RVQ), [6] studied the sum rate performance of RS as a function of the power splitting ratio $\rho$, which indicates the fraction of the total power allocated to the private messages. The optimal value of $\rho$ that maximizes the sum rate is determined as a function of the CSIT error, which is computable given the SNR, the number of transmit antennas $M$ and the number of feedback bits $B$. Those three parameters provide the necessary long-term information to the RS transceiver design.

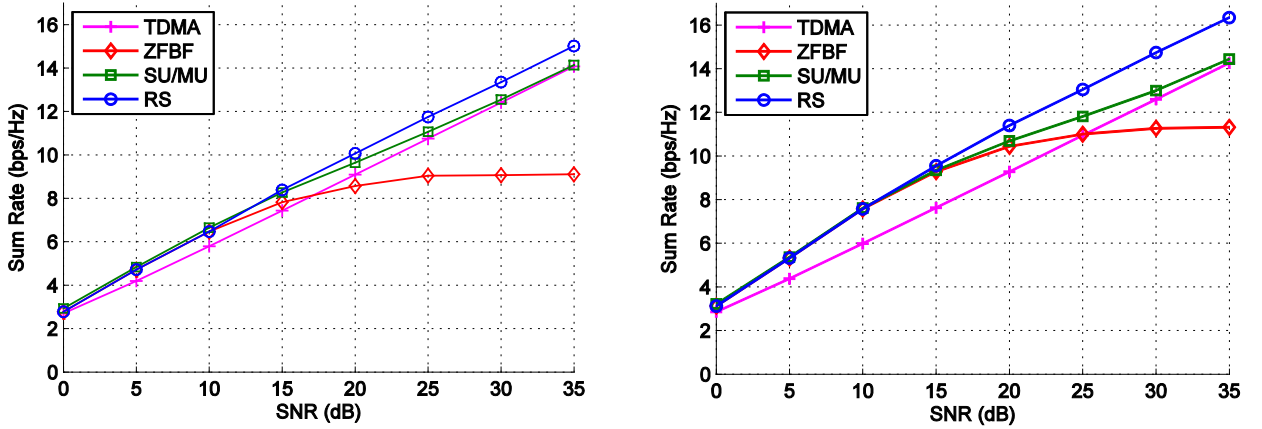

Figure 3. RS vs conventional schemes, *M*=4, *K*=2, *B*=10 (left), 15 (right).

Figure 3 illustrates the sum rate performance achieved with the RS approach when the number of feedback bits $B$ does not scale with SNR (equivalent to the case $\alpha = 0$). Performance of three conventional approaches are also displayed, namely the single-user mode 'TDMA', the multiuser mode 'ZFBF' and the single-user/multi-user mode switching 'SU/MU'. SU/MU dynamically switches between ZFBF and TDMA to maximize the sum rate. Four transmit antennas and two users are assumed. The precoders for the two private messages in RS are designed using ZFBF and allocated a fraction of the total power that is uniformly split among them, the remaining power being allocated to a common message. At low SNR, since the system is noise limited, the RS approach becomes ZFBF, i.e., $\rho = 1$. As the SNR increases, the power allocated to the common message increases, i.e., $\rho < 1$. At high SNR, the sum rate achieved by ZFBF saturates due to the interference limited-behavior created by the inaccurate CSIT, while the sum rate achieved by RS keeps increasing with a DoF of 1 because a dominant part of the total power is allocated to the common message.

In a practical system like LTE-A, the saturation of the sum rate is avoided by performing SU/MU. As mentioned above, at high SNR, the sum rate achieved by RS is dominated by the common message. However, since the common message has to be decoded by both users, its rate is limited by the weakest user and is probably lower than the rate of the single message sent via SU/MU because SU/MU boils down to TDMA at high SNR. Despite of that, Figure 3 shows that the contribution of the rates of the private and common messages altogether leads to a higher sum rate than SU/MU. Such a rate gap increases when $B$ grows up from 10 to 15.

Moreover, the number of feedback bits required by RS to maintain a constant sum rate gap relative to ZFBF with perfect CSIT is characterized in [6]. By comparing with the feedback overheard required by ZFBF with RVQ, it is shown that a significant feedback overhead reduction is enabled by RS. Setting e.g. the constant sum rate gap to be 6bps/Hz with 4 transmit antennas, at a medium SNR of 15dB, RS requires 5 bits less than ZFBF with RVQ to achieve the same performance.

## 5. Transceiver Optimization

Although ZFBF strategies achieve the optimum DoF, they are generally sub-optimal in finite SNR regimes where non-asymptotic metrics are considered, e.g. the Mean Square Error (MSE), the Signal to Interference plus Noise Ratio (SINR), and the achievable rate. Optimum precoders with respect to such metrics strike a delicate balance between nulling the undesired interference and maximizing the desired power components at the receivers. Generally speaking, optimum precoders can be hardly found in closed-forms, and obtaining them requires solving sophisticated optimization problems. The formulation of such problems strongly relies on the CSIT error model, which varies according the considered setup. For example, the BS may have access to some statistical properties of the CSIT error which can be employed to formulate average-based or outage-based problems. On the other hand, when the BS can only bound the CSIT error within some known uncertainty region, the optimization problem is formulated in terms of the worst-case performance. Another determining factor is the design's objective and constraints. For example, we may have a power constrained transmission with the objective of maximizing the sum-rate targeting the overall system performance, or the minimum rate among users to achieve a form of fairness. Alternatively, the design may also be Quality of Service (QoS) constrained with the objective of minimizing the transmission power required to achieve prescribed user rates.

One common feature in all RS optimization problems is the embedded sum-rate expressions. In particular, each user's achievable rate writes as the sum of two terms corresponding to the rates of the common and private parts of the message. Optimizing a two-part achievable rate for each user yields the optimum ratio in which messages should be divided for the given system setup. However, this also poses an optimization challenge since such sum-rate terms are non-convex and intractable in their original forms. This can be tackled through equivalent reformulations into special forms of Weighted MSE (WMSE) problems [7]. The domain of the original problem is extended by incorporating the receive filters and the MSE weights into the set of optimization variables. It can be shown that any optimum solution of the extended WMSE problem is also an optimum solution of the original rate problem. Moreover, the extended WMSE problem possesses a special structure that enables a solution using alternation optimization. However, due to the non-convexity of the original rate problem, global optimality cannot be guaranteed. However, extensive simulations have demonstrated that this approach is very efficient and achieves very good performances.

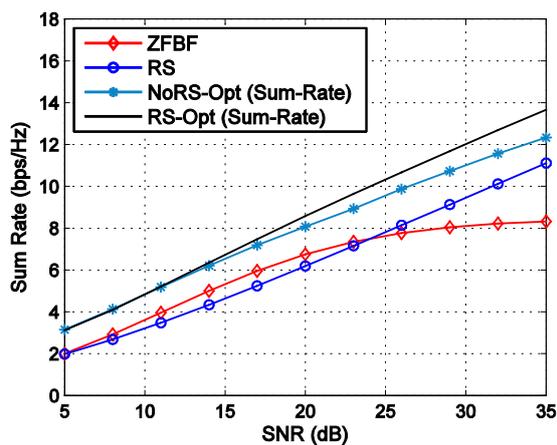

(a) $\alpha = 0$

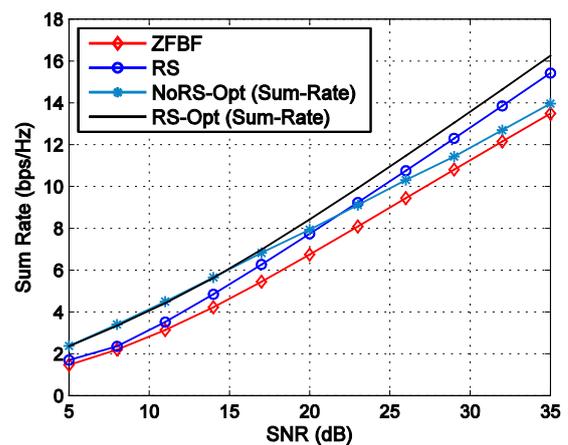

(b) $\alpha = 0.6$

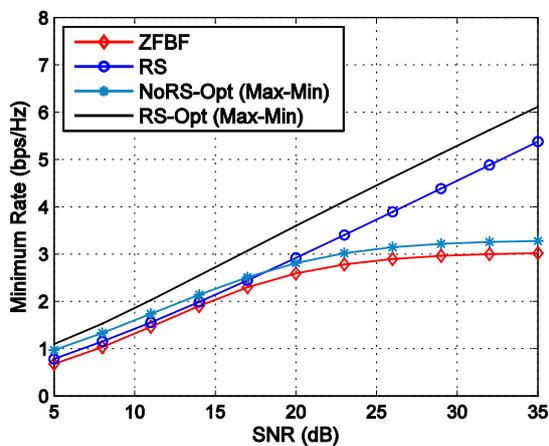

(c) $\alpha = 0$

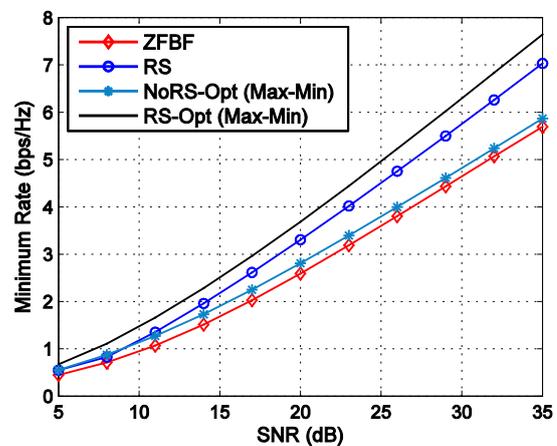

(d) $\alpha = 0.6$

Figure 4. Sum-rate and Minimum rate achieved with optimized and non-optimized precoders.

Figure 4 demonstrates the gains achieved by optimized precoders compared to simpler ZFBF based designs in the presence of i.i.d Gaussian CSIT errors with $\alpha = 0$ and $0.6$. Two design objectives are considered, maximizing the average sum rate and maximizing the minimum average rate. The superiority of optimized designs is clear for all cases. Further details on RS precoders optimization can be found in [8] for the sum-rate maximization and in [9] for the minimum rate maximization subject to a total transmit power constraint, and the power minimization under rate constraints.

## 6. Massive MIMO

When it comes to massive MIMO, a full dimensional channel estimate either requires an unaffordable feedback overhead in FDD or suffers from pilot contamination and antenna/RF miscalibration in TDD. Leveraging the rate, reliability and feedback overhead reduction benefits of RS in conventional MU-MIMO with imperfect CSIT, RS can be applied to tackle those Massive MIMO problems as demonstrated in [10]. However, since the common message has to be decoded by all co-scheduled users, its achievable rate degrades as the number of users increases. To retain the benefits of RS in such scenarios, a general RS framework, denoted as Hierarchical-Rate-Splitting (HRS), has been introduced in [10]. HRS exploits the knowledge of transmit correlation matrices to alleviate the CSIT requirement and transmits two kinds of common messages to mitigate the rate constraints of the common message.

To do so, users are clustered into groups according to the similarity of their transmit correlation matrices. Then, a two-tier downlink precoder, that is reminiscent of the dual-codebook structure of LTE-A [2], is adopted: the outer precoder controls inter-group interference based on long-term CSIT while the inner precoder controls intra-group interference based on short-term effective channel. Due to imperfect grouping and instantaneous CSIT, residual inter-/intra-group interference remain the limiting factors of the system performance. To overcome this problem, the philosophy of RS is generalized into HRS which consists of an outer RS and an inner RS, as illustrated in Figure 5 (left). By treating each group as a single user, an outer RS tackles the inter-group interference by packing part of the users' messages into a common message $s_0$ that can be decoded by all users. Likewise, an inner RS copes with the intra-group interference by packing part of the messages intended to users in that group (say $g$) into a common message $s_{g,0}$ that can be decoded by group $g$'s users. The common messages are transmitted along the private messages in a superimposed manner. At the receiver side, each user sequentially decodes $s_0$ and $s_{g,0}$ of its corresponding group, and removes them from the received signal by SIC. Then, the private message of each user can be independently decoded by treating all other private messages as noise. When the inter-group interference is negligible, HRS becomes a set of parallel RS in each group. By contrast, when the inter-group interference is the dominant degrading factor, HRS boils down to be RS.

The performance gain of HRS over RS and conventional approaches is illustrated in Figure 5 (right) for $M = 100$ and SNR=30dB in a typical scenario where the inter-group interference is negligible.

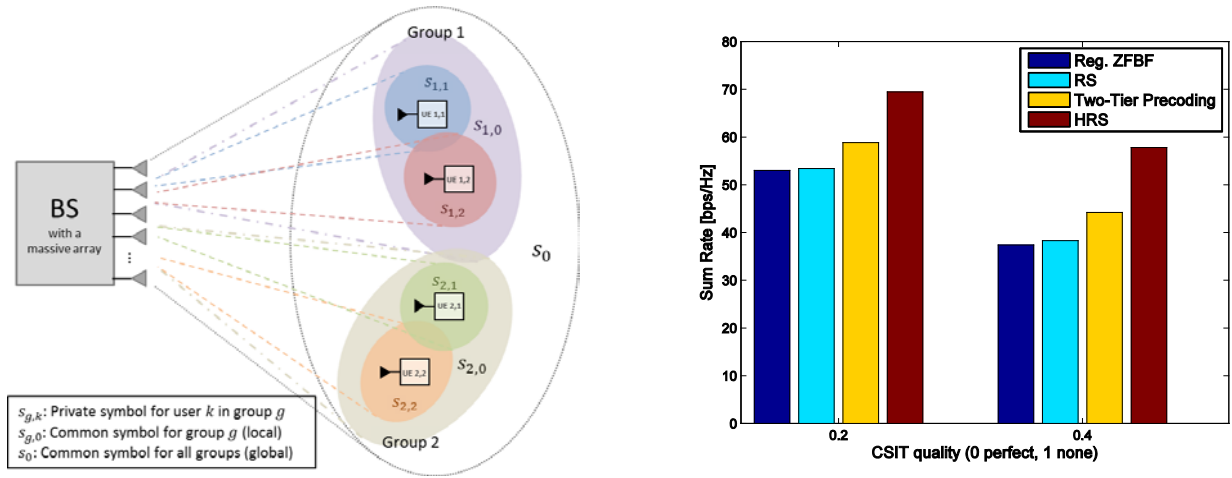

Figure 5. HRS for Massive MIMO: Architecture (left) and Performance (right)

## 7. Multi-Cell Coordination

In LTE-A, CoMP has been included in the specification as a technique to deal with the inter-cell interferences. However, CoMP has only partially convinced industry in 3GPP. The large disparity of performance results on CoMP [1] also highlights the lack of reliability and the high sensitivity of such techniques. Imperfect CSIT among the coordinated/cooperative transmitters is the major issue that impacts CoMP system throughput. RS can be used to enhance the system performance in the multi-cell scenario in the presence of imperfect CSIT.

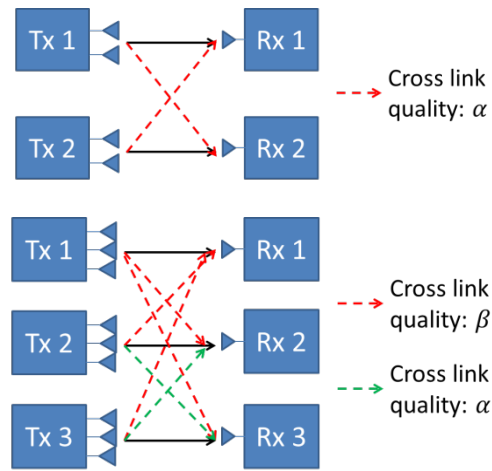

Figure 6. Multi-cell coordination in a two-cell scenario with symmetric CSIT qualities (above) and three-cell scenario with a certain CSIT quality topology (below).

Let us first consider the two-cell scenario illustrated in Figure 6 (above), where each transmitter is equipped with two antennas and each user has a single antenna. The two transmitters share the CSI of the two users, but not the user data. Since the CSIT qualities of the cross links are $\alpha$, the sum DoF achieved with ZFBF is $2\alpha$. However, with RS, each transmitter delivers the ZFBF-precoded private message using a fraction of the power while one transmitter sends a common message using the remaining power. The sum DoF is enhanced to $1 + \alpha$ [11].

In the three-cell scenario, since the interferences overheard by a single user come from two different cross links, transmitting a common message to be decoded by all users may not properly cope with the interference between a pair of two users. To overcome this, as an evolution of RS, a Topological RS (TRS) is introduced in [12]. It consists of a multi-layer structure (somewhat reminiscent of the HRS strategy) and transmits multiple common messages according to the CSIT quality topology. For instance, let us focus on the scenario illustrated in Figure 6 (below), where each transmitter has three antennas and each user has a single antenna. User 2 and user 3 are grouped and user 1 alone forms another group, since user 2 and user 3 have identical intra-group CSIT qualities, i.e., $\alpha$, which is smaller than the inter-group CSIT qualities, i.e., $\beta$. Then, similarly to HRS, the TRS transmitted signal is formed by a two-layer RS. The outer layer tackles the inter-group interferences by transmitting a system common message to be decoded by all users, private message of user 1, and private messages for the group formed by user 2 and user 3. The private messages for the group formed by user 2 and user 3 are referred to as the inner layer. It comprises two private messages plus a group common message in order to deal with the intra-group interference. By performing SIC, user 2 and user 3 decode the system common message, group common message and their desired private messages, sequentially. On the other hand, user 1 only decodes the system common message and its desired private message.

More details on TRS for the general K-cell scenario with arbitrary CSIT quality topologies can be found in [12].

## 8. Rate-Splitting in LTE Evolution

(H/T)RS is a generalized strategy that incorporates conventional SU/MU-MIMO and CoMP as special cases, i.e. whenever the power allocated to the common message(s) is set to 0. This enables a more general form of mode switching in LTE Evolution where switching can be operated between SU, conventional MU and RS depending on the SNR and the CSIT quality.

The introduction of RS in LTE Evolution would have various impacts on the standardization efforts. A new transmission mode indicator, in the form of a DCI format, is needed to tell the user the proper transmission mode and the relevant information required for demodulation. The receiver also needs to be informed about the type of messages (common/private), the number of messages, the modulation and coding scheme of all common/private message intended for the user, information about whether common message is intended for the user or not, the transmit power of each message, etc. In the uplink, RS also has an impact on the CSI feedback mechanisms and signaling. RS requires the knowledge about the CSIT accuracy in order to properly allocate the power to the common and private messages. This could be for instance computed by a UE and reported back to the BS. Based on the collection of those CSIT accuracies from all users in all subbands, the BS performs user scheduling and decides upon the appropriate transmission strategy. RS also has an impact on the fundamental CSI feedback mechanisms on PUCCH and PUSCH, i.e. in what time and frequency resource, the CSI of a given UE is reported. It is indeed shown in [13] that some CSIT patterns lead to a higher DoF than others.

## 9. Conclusions and Future Challenges

Contrary to the LTE-A design that relies on private message transmissions (only motivated in the presence of perfect CSIT), this paper introduces a promising Rate-Splitting (RS) strategy relying on the transmission of common and private messages suitable for the realistic scenario of imperfect CSIT. The paper highlights the benefits of RS in terms of spectral and energy efficiencies, reliability and CSI feedback overhead reduction over conventional strategies as used in LTE-A.

RS has the potential to fundamentally change the design of the PHY and Lower MAC Layers of LTE Evolution. We here touched upon a few scenarios, briefly covering some aspects of MU-MIMO, Massive MIMO and multi-point coordination. RS is a golden mine of research problems for academia and of standard specification issues for industry. Just to name a few, RS has or is likely to have a significant impact on transmission schemes/modes, CSI feedback mechanisms, MIMO receiver implementation, user pairing, user and message scheduling, multi-carrier transmissions and novel 5G waveforms, spectral vs energy efficiency trade-off, highly reliable communications, NOMA/MUST, Massive MIMO, higher frequency bands operation (e.g. millimeter-wave), coordination/cooperation among distributed antennas in homogeneous and heterogeneous network deployments, interference alignment and network MIMO, relay channel, superposition of multicast and unicast messages and networks relying on pro-active caching.

**References**
[1] 3GPP TR 36.819 v11.0.0, "Coordinated multi-point operation for LTE physical layer aspects," Sept. 2011.
[2] C. Lim et al., "Recent Trend of Multiuser MIMO in LTE-Advanced," IEEE Comm. Mag., vol. 51, no. 3, pp. 127-135, Mar. 2013.
[3] B. Clerckx et al., "A Practical Cooperative Multicell MIMO-OFDMA Network based on Rank Coordination," IEEE Trans. on Wireless Comm. vol. 12, no. 4, pp. 1481-1491, Apr. 2013.
[4] A. El Gamal et al., *Network information theory*. Cambridge university press, 2011.
[5] S. Yang et al., "Degrees of freedom of time correlated MISO broadcast channel with delayed CSIT," IEEE Trans. Inf. Theory, vol. 59, no. 1, pp. 315–328, Jan. 2013.
[6] C. Hao et al., "Rate analysis of two-receiver MISO broadcast channel with finite rate feedback: A rate-splitting approach", IEEE Trans. on Comm., vol. 63, no. 9, pp. 3232-3246, Sept. 2015.
[7] S. Christensen et al., "Weighted sum-rate maximization using weighted MMSE for MIMO BC beamforming design," IEEE Trans. on Wireless Comm., vol. 7, no. 12, pp. 4792–4799, Dec. 2008.
[8] H. Joudeh et al., "Sum-Rate Maximization for Linearly Precoded Downlink Multiuser MISO Systems with Partial CSIT: A Rate-Splitting Approach," arXiv:1602.09028, Feb. 2016.
[9] H. Joudeh et al., "Robust Transmission in Downlink Multiuser MISO Systems: A Rate-Splitting Approach," arXiv:1602.04345, Feb. 2016.
[10] M. Dai, et al., "A Rate Splitting Strategy for Massive MIMO with Imperfect CSIT," IEEE IEEE Trans. on Wireless Comm., accepted for publication, arXiv:1512.07221, Dec. 2015.
[11] X. Yi et al., "On the DoF of the multiple-antenna time correlated interference channel with delayed CSIT," in ASILOMAR 2012.
[12] C. Hao et al., "MISO Networks with Imperfect CSIT: A Topological Rate Splitting Approach", arXiv:1602.03768, Feb. 2016.
[13] B. Rassouli et al., "DoF Analysis of the MIMO Broadcast Channel with Alternating/Hybrid CSIT," IEEE Trans. on Info Theory, vol. 62, no. 3, pp. 1312-1325, Mar. 2016. .